# Phase Transitions Microscopic Model

V.A.Stepanov

SSC RF – Institute of Physics and Power Engineering,
249033 Obninsk, Russia e-mail: stepanov@ippe.obninsk.ru

The microscopic model in which nodes interacting with each other are statistical systems is introduced. The nodes conditions are connected with a string of distinct microscopic configurations and depend on external parameters (pressure and temperature). In this model the consequent description of first- and second-order phase transitions is carried out and their microscopic level distinctions are analyzed. It is shown that first-order transitions occur when the configuration entropy change at nodes under transitions from dipole-active (low-symmetric) state to symmetric one is more than ln4. Otherwise the second-order transitions take place.

The merit of microscopic models of phase transitions is the opportunity to get evident explicit of dependence of non-equilibrium thermodynamic potential from order parameter $F(h)$ which may be used to analyze a correlation of microscopic parameters of systems with phase transitions types. The most well known are the Ising model and its various modifications in which dipole states $\mu=\pm 1/2$ or $\mu=0,\pm 1$ occur at each N nodes and the lattice interaction energy is appeared as

$$E_N = -J \sum_{i,j} m_i m_j + \sum_i f(m_i), \quad (1)$$

where J – dipole-dipole interaction energy, and the second term depends on nodes energy states spectrum. Unlike the simple Ising model in which only second-order phase transitions can occur the type (1) models allow to obtain such states of systems that correspond both single and two-phased areas on the phase diagrams [1]. However, to describe, for example, first-order phase transitions the interdependence of microscopic parameters is to be assumed which is physically unfounded. Thus the consequent description first-order phase transitions appears to be impossible, not to mention the analysis of distinctions between first- and second-order phase transitions on microscopic level. It is explained by the fact that models (1) lack parameters, related to the structure, to the short order and to the various state probabilities at nodes. In fact, first-order phase transitions, as a rule, are connected with essential structure modifications. In the first-order phase transition point a change of system internal energy takes place along with entropy change, reflecting a probability change of microscopic states during the transition from one phase to another or a change in number of means through which a physical quantity connected with phase transition in the system is realized.

In the expression for energy of the system the configuration entropy change during dipole transitions µ=0, ±1 at nodes is to be taken into consideration.

$$E_N = -J \sum_{i,j} m_i m_j - Ts \sum_i |m_i| \qquad (2)$$

s means difference configuration entropies of states µ=±1 and µ=0 at a node. The conception of node is interpreted in wide sense. For example, it may be as either single atom, or a group of atoms, various configurations of which correspond to a certain value of µ. In mean field approximation ($<\mu_i> = \mathbf{h}$) non-equilibrium potential for (2) will be written as

$$F = \frac{qJ\mathbf{h}^2}{2} - \frac{1}{\mathbf{b}} \ln[1 + 2 \cdot \exp(s) \cdot ch(qJ\mathbf{hb})], \qquad (3)$$

where $ß^{-1} = T$ is temperature in energetic units. The configuration entropy meaning of s is discovered in microscopic description which in mean field approximation allows us to set the configuration of non-equilibrium potential by the appointment of the state probabilities with different µ at given node:

$$\frac{1}{qJ} \frac{\partial F}{\partial \mathbf{h}} = \mathbf{h} - \sum_{\mathbf{m}} \mathbf{m} \cdot p_{\mathbf{m}} , \qquad (4)$$

where $p_\mu$ is state probability µ at node, q is number of neighbor nodes.

Under microscopic kinetic approach state probabilities at a node are determined on the basis of account of transition probability between states. For states µ=0,±1:

$$\frac{1}{qJ}\frac{\partial F}{\partial h}=h-\frac{\dfrac{v(0,1)}{v(1,0)}-\dfrac{v(0,-1)}{v(-1,0)}}{1+\dfrac{v(0,1)}{v(1,0)}+\dfrac{v(0,-1)}{v(-1,0)}}, \qquad (5)$$

where probabilities (frequencies) ratio depends on a variety of energies between states of µ and on level number $N_\mu$ under these states:

$$\frac{v(0,1)}{v(1,0)}=\frac{N_1}{N_0}\exp(-bE_1), \qquad \frac{v(0,-1)}{v(-1,0)}=\frac{N_1}{N_0}\exp(-bE_{-1}) \qquad (6)$$

If the state node spectrum is proposed as $E_1=-qJ\boldsymbol{h}$, $E_0=0$, $E_{-1}=qJ\boldsymbol{h}$ then from (5) and (6) one shall get:

$$F = \frac{qJ\boldsymbol{h}^2}{2} - \frac{1}{b}\ln\left[1+2\cdot\frac{N_1}{N_0}ch(qJ\boldsymbol{h}b)\right], \qquad (7)$$

those in comparison with (3) leads to $s=\ln\dfrac{N_1}{N_0}$. Configuration numbers $N_\mu$ are connected with nodes by internal structure and depend on their symmetry in µ states. The lattice systems are possible, for example, in crystals where dipole states $\mu^k$, conforming with various physical properties of $<\mu^k>$, either at the same nodes or at another ones, are realized. Symmetry and structure parameters, magnetization, electric polarization can represent these physical properties. In cases when $N_\mu$ depends on realizations of other $\mu^k$, an interaction of order parameters $\boldsymbol{h}$ and $\boldsymbol{h}^k$ occurs by configuration entropy s.

Figure 1 shows a phase diagram calculating for non-equilibrium potential (3) in the coordinates T/qJ – exp(-s). In diagram below the line **ab** of the stability boundary of the disordered phase ($\boldsymbol{h}$=0) only phase with $\boldsymbol{h}\neq 0$ is stable. Line **acd** is the boundary of the order phase stability. Above this line only phase with $\boldsymbol{h}$=0 is stable. The region **dcb** is a two-phase one where non-equilibrium potential has two symmetric minima at $\boldsymbol{h}\neq 0$ and minimum at $\boldsymbol{h}$=0. Along line **ce** of first-order phase transitions the values of F in minima are equal. Critical point **c** has coordinates (4,1/3) which can be founded under condition when $2^{nd}$ and $4^{th}$ derivates of non-equilibrium potential are equal zero at $\boldsymbol{h}$=0.

The analysis of state diagram allows to draw a conclusion that in systems with dipole-dipole interaction the first-order transitions take place in that case when ratio of the realization number of dipole-inactive state to the configuration number of dipole-active state (entropy factor exp(-s)) at nodes is more than 4. Otherwise, the second-order phase transitions occur. First-order phase transitions take place when a change of configuration entropy at nodes under transition from dipole-active (low-symmetric) state to symmetric one is more than ln4.

The first-order phase transition heat $\Delta H$ is equal to a difference of mean values of entropy terms for high- and low-symmetric phases in expression (2):

$$\Delta H = -\frac{Ts}{n}(\langle|m|\rangle_{h=h_0} - \langle|m|\rangle_{h=0}), \tag{8}$$

where averaging of $|m|$ is carried out for asymmetric with $h=h_0$ and symmetric with $h=0$ phases at a phase transition point, n is a number of atoms at a node. Equation (8) provides an expression, connecting a microscopic configuration entropy s and with first-order transition heat:

$$\frac{\Delta H}{T}n = r(s), \tag{9}$$

where the function

$$r(s) = \ln(\exp(-s))\left[\frac{2ch(qJh_0 b_0)}{\exp(-s)+2ch(qJh_0 b_0)} - \frac{2}{\exp(-s)+2}\right]$$

one calculates numerically.

For structural phase transitions in solids the entropy factor value exp(-s) related to a change of symmetry element number is usually small and r(s)<1. For this reason the meaning $\Delta H/T$ is much less 1. It was proved by experimental data. Entropy factor exp(-s) estimation from (9) for various phase transitions in solids gives its values in range from 4 to 4.5. When solids are melted the microscopic configuration number of node symmetric states exceeds greatly the number of low-symmetric ones, connected with symmetry crystal configurations at nodes if it is proposed that approach mentioned above to melting-crystallization phase transitions to be corrected. For example, when simple metal is melted $\Delta H/T\sim 1$-1.3. When one purposes n~2 (a number of atoms in unit cell) from (9) one can receive r(s)~2-2.6 and $N_0/N_1$~15-20. Final set of states under μ=0 assumes the existence of short range ordering in liquid.

In case when during node transition from state μ=±1 to the one with μ=0 there is a volume change on value ΔV, under external pressure action P the node state spectrum is $E_1$=-qJ**h**-PΔV, $E_0$=0, $E_{-1}$=qJ**h**-PΔV. The corresponding expression for non-equilibrium potential from (5) and (6) is

$$F = \frac{qJh^2}{2} - \frac{1}{b}\ln[1 + 2 \cdot \exp(s + P\Delta Vb) \cdot ch(qJhb)] \qquad (10)$$

Phase diagrams in dimensionless temperature T/qJ and pressure p=PΔV/qJ coordinates for systems with various entropy factor meanings are shown in Fig.2. Diagrams were designed for positive and negative meanings of ΔV. If the phase transitions are accompanied by positive voluminous effect than the pressure increase corresponds to the right part of the diagram with p>0. If the symmetric phase has lower volume the pressure increase corresponds to the left part of the diagram with p<0. On the diagram the region above line **acb** conforms to a symmetric phase with **h**=0, below line **0cb** – to low-symmetric phase with **h**≠0. Region **-1c0** is a two-phased one. Critical point (**c, c', c''**) has coordinates T/qJ=1/3, p=(-s-ln4)/3. The phase diagram gives clear vision that for first-order phase transition (exp(-s)>4) with positive voluminous effect the transition temperature grows along with a pressure increase, so when pressure is $P_c$=(-s-ln4)$T_c$/ΔV the phase transition transforms into a second-order transition.

If first-order phase transitions have negative voluminous effect (for example, ice – water, graphite – diamond, etc.) than a critical point in phase diagram is absent and under high pressure and low temperature a symmetric and a low-symmetric phases coexist. For second-order phase transitions (exp(-s)<4) the transformation to a first-order phase transition with a pressure increase is possible only when a voluminous effect of transition is negative.

The next expression of energy of system is corresponding to non-equilibrium thermodynamic potential (10):

$$E_N = -J\sum_{i,j} m_i m_j - [P\Delta V + Ts] \cdot \sum_i |m_i| . \qquad (11)$$

In this expression each of the interacting nodes through dipole-dipole mechanism is a statistical system which state conditions depend on microscopic parameters s and $\Delta V$ and external parameters of pressure and temperature.

**Literature.**

[1] Plaksin O.A., Stepanov V.A. Phase Transitions, 1992, Vol.40, pp. 105-112.

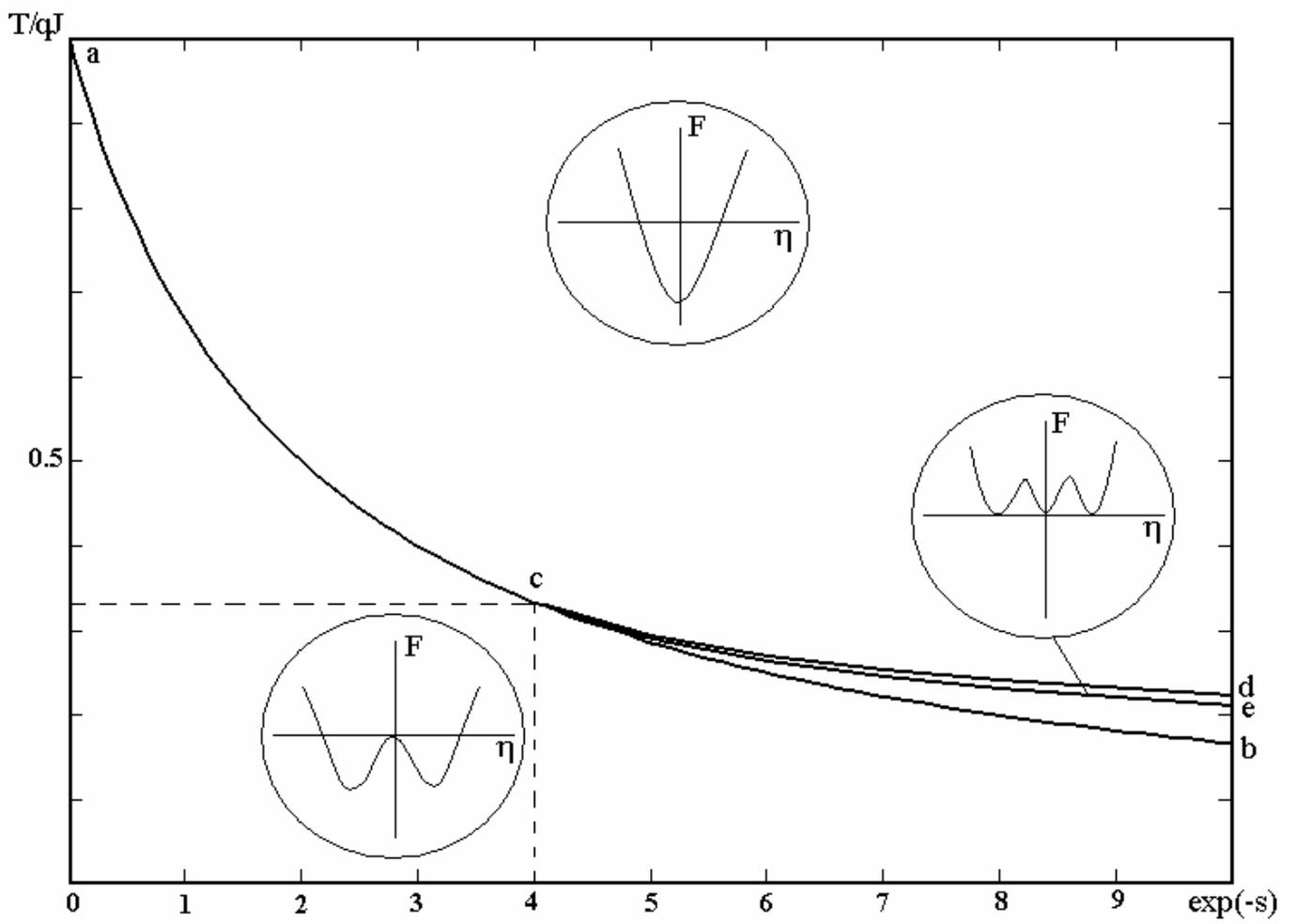

Fig.1. The phase diagram in the coordinates $T/qJ$ – $\exp(-s)$. The inserts give the types of the non-equilibrium potential for different regions.

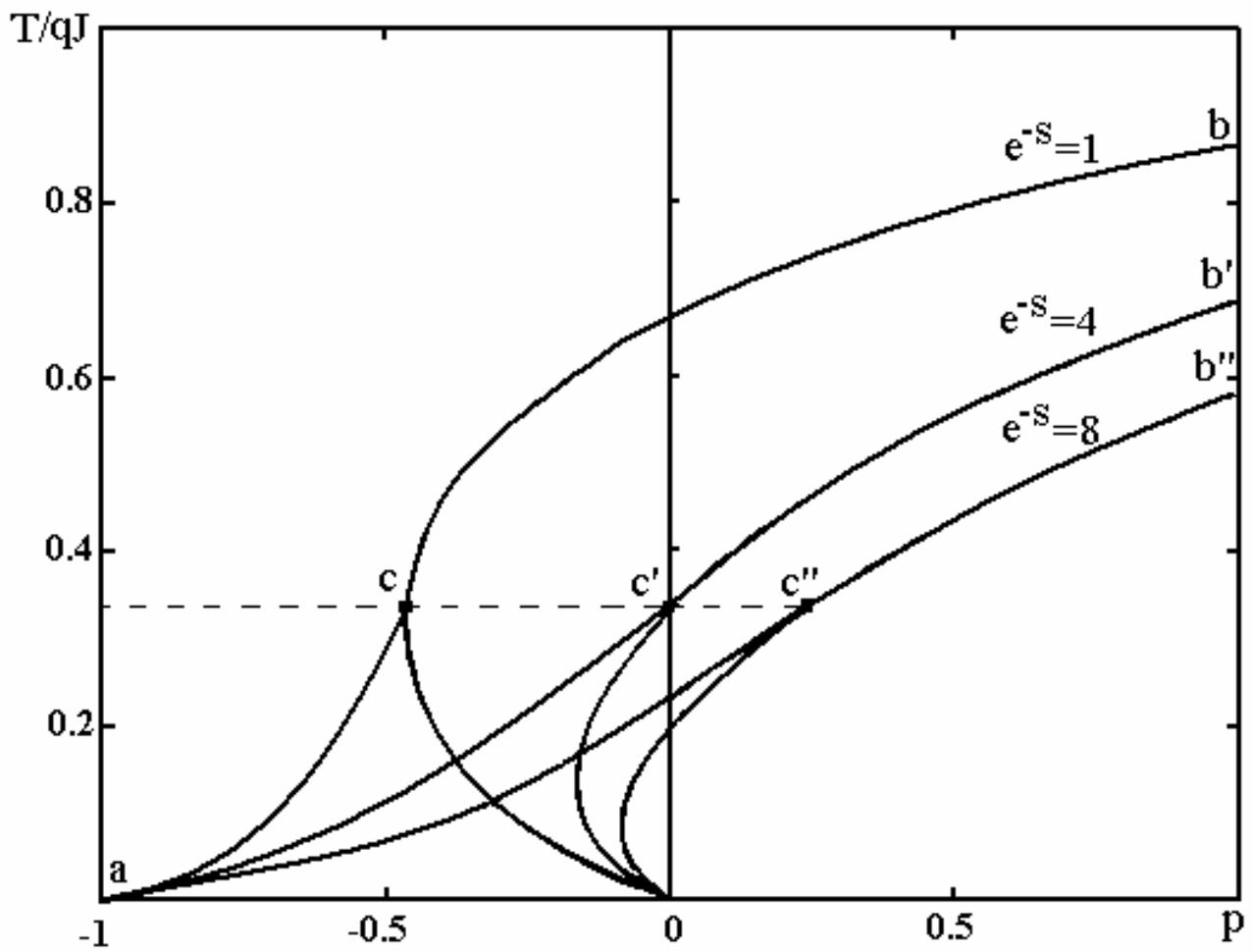

Fig.2. The phase diagram in the coordinates temperature (T/qJ) – pressure (p=PΔV/qJ) for systems with various entropy factor values (exp(-s)).